\documentclass{ws-rv9x6}
\usepackage{subfigure}
\usepackage{ws-rv-thm}
\usepackage{ws-rv-van}
\usepackage{mathtools}

\makeindex

\begin{document}
	
\chapter[Loop braid groups and integrable models]{Loop braid groups and integrable models\label{chap: loopbraid}}

\author[Pramod Padmanabhan \& Abhishek Chowdhury]{Pramod Padmanabhan, Abhishek Chowdhury}

\address{{\it School of Basic Sciences,\\ Indian Institute of Technology, Bhubaneswar, India} \\
	ppadmana, achowdhury@iitbbs.ac.in}

\begin{abstract}
	Loop braid groups characterize the exchange of extended objects, namely loops, in three dimensional space generalizing the notion of braid groups that describe the exchange of point particles in two dimensional space. Their interest in physics stems from the fact that they capture anyonic statistics in three dimensions which is otherwise known to only exist for point particles on the plane. Here we explore another direction where the algebraic relations of the loop braid groups can play a role -- quantum integrable models. We show that the {\it symmetric loop braid group} can naturally give rise to solutions of the Yang--Baxter equation, proving the integrability of certain models through the RTT relation. For certain representations of the symmetric loop braid group we obtain integrable deformations of the $XXX$--, $XXZ$-- and $XYZ$--spin chains.
\end{abstract}

\body

\section{Introduction}
\label{sec:Introduction}
Two dimensional spaces offer more possibilities for point particles--{\it anyons}--that obey exotic statistics, unlike their three dimensional counterparts that appear as either bosons or fermions. However this narrative fails for {\it extended objects} in higher dimensions giving rise to novel statistical properties and spin--statistics relations~\cite{bal1, sor1, sor2, bal2, fried}. While the two dimensional anyons are seen as representations of the braid group, the statistics group of extended objects in higher dimensional spaces are the {\it motion groups}~\cite{motion}. Anyons have received renewed attention in recent years due to their potential applications in {\it topological quantum computation}~\cite{tqc}.

Though these questions address some important aspects of fundamental physics, the physicist's interest in the braid group also extends to the study of {\it quantum integrable models} which form the basis of the method of {\it algebraic Bethe ansatz}~\cite{slav, korepin, jimbo}. The method of {\it Baxterization}~\cite{jones} helps us obtain $R$--matrices that solve the {\it spectral parameter}--dependent {\it Yang--Baxter equation} (YBE) from braid group generators. In this paper we show the construction of $R$--matrices using the generators of a quotient of the {\it loop braid group}--the {\it symmetric loop braid group}~\cite{loop1, loop2, loop3}, that study the exchange of loops ($S^1$) in three dimensions. We will see that they naturally give rise to solutions of the YBE and hence open the possibility for obtaining new integrable models.

The contents of this article are as follows : We review the construction of local integrable models using the $R$--matrix method in Sec.~\ref{sec:Rmatrixintegrability}. We then move on to describing the algebraic relations of the loop braid group in Sec.~\ref{sec:loopbraidgroup} and use these to construct YBE solutions in Sec.~\ref{sec:LBGRmatrices}. By choosing appropriate representations of the symmetric loop braid group we find associated integrable models in Sec.~\ref{sec:LBGspinchains}. We end with an outlook in Sec.~\ref{sec:Outlook}. \\

{\it This article is to honor the immense contribution of A. P. Balachandran to theoretical physics and its community on the occasion of his $85^{th}$ birthday. Bal, as he is fondly known among his friends, has been a continuous source of inspiration for several generations of physicists including us and this work is one such example coming from his school of thought.} 

\section{Integrability {\it via} $R$--matrices}
\label{sec:Rmatrixintegrability}
In 1931, Hans Bethe proposed an original method (\textit{the coordinate Bethe ansatz}) to arrive at the spectrum of the quantum Heisenberg spin chain which later on in 1970--80's gave rise to a new algebraic approach (\textit{the algebraic Bethe ansatz}) to study a wide class of quantum systems in various dimensions. It was found that the same algebra of operators had different representations which solved many quantum models with completely different physical interpretations. 

The algebraic Bethe ansatz (ABA) provides a mathematical reasoning for the integrability of these myriad quantum systems by arriving at the existence of an infinite dimensional symmetry. 
Here, we will focus on closed spin chains with $N$ sites and periodic boundary conditions, whose Hilbert space has the form
\begin{equation}
	\mathcal{H}=\bigotimes_{n=1}^{N} \mathcal{H}_{n}, 
\end{equation}
where $\mathcal{H}_n$ is the Hilbert space at the $n^{th}$ site of the chain.

The fundamental object of ABA is a generating object called the \textit{Lax operator} or the \textit{monodromy matrix}, $T(u)$. In order to define it, we have to introduce a auxiliary space $V$ (which we take as $\mathbb{C}^2$) in addition to the Hilbert space $\mathcal{H}$ of the quantum system. We also have to introduce a continuous complex parameter $u$, the \textit{spectral parameter} which allow us to recover the integrals of motions as coefficients of a series expansion in $u$. The auxiliary space $V$ on the other hand is essential to show that the integrals of motion commute. The Lax operator acts on $\mathcal{H}\otimes V$ i.e. the matrix elements $T^{ij}$ are operators acting on the Hilbert space $\mathcal{H}$:
\begin{equation}
	T(u)=\left(\begin{array}{ll}
		A(u) & B(u) \\
		C(u) & D(u)
	\end{array}\right).
\end{equation}
They are assumed to follow the commutation relations (RTT) in the space $\mathcal{H}\otimes V_1 \otimes V_2$ : 
\begin{equation}
	\label{eq:RTT}
	R_{12}(u-v) T_{1}(u) T_{2}(v)=T_{2}(v) T_{1}(u) R_{12}(u-v),
\end{equation}
where the \textit{R--matrix} $R_{12}(u)$ acts on $V_1\otimes V_2$ and are solutions to the YBE acting on the tensor product of auxiliary spaces $V_1\otimes V_2 \otimes V_3$ :
\begin{equation}
	R_{12}\left(u\right) R_{13}\left(u +v\right) R_{23}\left(v\right)=R_{23}\left(v\right) R_{13}\left(u+v\right) R_{12}\left(u\right).
\end{equation}
It is straightforward  to verify that $\displaystyle R(u)=u\,\mathbb{I} + c\, s$ (we have the freedom to multiply by an arbitrary function) where $c$ is a constant and $s$ (defined in Sec.~\ref{sec:loopbraidgroup}) is the permutation matrix, solves the above YBE. For this $R$--matrix, Eq.~(\ref{eq:RTT}) leads to the following commutation relations:
\begin{equation}
	\left[T^{i j}(u), T^{k l}(v)\right]=\frac{c}{u-v}\left(T^{k j}(v) T^{i l}(u)-T^{k j}(u) T^{i l}(v)\right),
\end{equation}
which gives the $ABCD$ algebra crucial to generating the spectrum of the given quantum system. 


From the RTT relation Eq.~(\ref{eq:RTT}), multiplying by $\displaystyle R^{-1}_{12}(u-v)$ from the left, taking trace w.r.t. the space $V_1\otimes V_2$ and using the cyclic permutations in the trace, we get
$\displaystyle \left [\mathcal{T}(u),  \mathcal{T}(v)\right]=0$. It is only natural to work with the \textit{transfer matrix}, $\displaystyle \mathcal{T}(u)=\operatorname{tr} T(u)=A(u)+D(u)$ as it satisfies the periodic boundary condition of one--dimensional quantum chains. Expanding $\mathcal{T}(u)$ into a power series over $u$ centered around some point $u_0$ gives
\begin{equation}
	\mathcal{T}(u)=\sum_{k}\left(u-u_{0}\right)^{k} I_{k},
\end{equation}
which leads to a set of commuting operators in the Hilbert space $\mathcal{H}$
\begin{equation}
	\left[I_{k}, I_{n}\right]=0, \quad \forall~ k, n.
\end{equation}
If we now set one of these $I_k$ to be the Hamiltonian of some quantum system, then we have an infinite set of integrals of motion (for a finite chain of size $N$, the expansion will get truncated at $k=N$), i.e. an integrable system. Before discussing the  Hamiltonians, we note two important points:
\begin{itemlist}
	\item In order to ensure locality of the operators $I_k$, the correct choice(s) of decomposition point $u_0$ is crucial.
	\item Instead of expanding the operator $\mathcal{T}(u)$ into a series, we could have expanded a function of this operator (almost always the case).
\end{itemlist}
In the section below, we will demonstrate this with the concrete example of the $XXX$--spin chain.

\subsection*{Local Hamiltonians}
\label{subsec:XXXchain}
Now we show one way of obtaining a local operator from the transfer matrix, $\mathcal{T}(u)$ through the expression
\begin{equation}
	H(u_0) = \frac{d\mathcal{T}(u)}{du}\mathcal{T}^{-1}(u)\Bigg|_{u=u_0},
\end{equation}
where the derivative is taken at some point $u=u_0$. We give the final answer in terms of a general invertible $R$--matrix and show that the $XXX$--spin chain can be obtained for a particular choice of the $R$--matrix.

A particular solution of the RTT relation, Eq.~(\ref{eq:RTT}) is the $R$--matrix itself and hence using this we obtain the transfer matrix, $\mathcal{T}(u)$ as 
\begin{equation}
	\mathcal{T}(u) = \operatorname{tr}_0\left[R_{0N}(u)\cdots R_{01}(u)\right],
\end{equation}
where the entries of the $R$--matrix act on the Hilbert space,  $\mathcal{H}=\otimes_{n=1}^N~\mathbb{C}^2_n$, the $N$--site spin chain. The auxiliary space is denoted by $V_0$. The derivative of $\mathcal{T}(u)$ is evaluated as follows,
\begin{eqnarray}
	\frac{d\mathcal{T}(u)}{du}\Bigg|_{u_0} & = & \sum\limits_{k=1}^N~\operatorname{tr}_0\left[R_{0N}(u_0)\cdots R'_{0k}(u_0) \cdots R_{01}(u_0)\right] \nonumber \\
	& = & \sum\limits_{k=1}^N~\operatorname{tr}_0\left[R'_{0k}\cdots R_{01}R_{0N}\cdots R_{0k+1}s_{0, k+1}^2\right] \nonumber \\
	& = & \sum\limits_{k=1}^N\operatorname{tr}_0\left[R'_{k+1,k}\cdots R_{k+1,1}R_{k+1,N}\cdots R_{k+1,k+2}R_{k+1, 0}\right] \nonumber \\
	& = & \sum\limits_{k=1}^N~\left[R'_{k+1,k}\cdots R_{k+1,1}R_{k+1,N}\cdots R_{k+1,k+2}\right],
\end{eqnarray}
where we have repeatedly used the cyclicity properties of the trace operation and we also take $\operatorname{tr}_0(R_{k+1,0}) \propto 1$. This is indeed the case in the examples we consider, nevertheless relaxing this assumption slightly modifies the computations and can still be handled in this procedure. In a similar fashion we find 
\begin{equation}
	\mathcal{T}(u)^{-1}\bigg|_{u=u_0}  = R^{-1}_{k+1,k+2}\cdots R^{-1}_{k+1, N} R^{-1}_{k+1,1}\cdots R^{-1}_{k+1,k}, 
\end{equation}
and multiplying this with the derivative of $\mathcal{T}(u)$ we obtain the Hamiltonian as
\begin{equation}\label{eq:localH}
	H(u_0) = \sum\limits_{k=1}^{N}~R'_{k+1,k}(u_0)R^{-1}_{k+1,k}(u_0).
\end{equation}
Note that though this operator is a sum of {\it local} terms, for a general $R$--matrix it need not be {\it self--adjoint}. 

On choosing $R(u) = u\,\mathbb{I} + c\,s$, we obtain the $XXX$--spin chain with the local Hamiltonian,

\begin{equation}
	H\left(\frac{c}{2}\right) = \frac{2}{3c}~\sum\limits_{k=1}^{N}~\left[X_{k+1}X_k + Y_{k+1}Y_k, + Z_{k+1}Z_k\right], 
\end{equation}
where we have used the representation of the permutation operator, $s$ (See Eq.~(\ref{eq:Permrep})). 

\section{Loop braid group}
\label{sec:loopbraidgroup}
The loop braid group, $LB_N$, is defined on $N$ sites just as the braid group. A convenient presentation is given by the relations of the generators, $\sigma_i$, $s_i$ with $i\in\{1,\cdots, N-1\}$. The braid generators, $\sigma_i$, satisfy the braid group relations,
\begin{align}
	\sigma_i\sigma_{i+1}\sigma_i & = \sigma_{i+1}\sigma_i\sigma_{i+1}, \tag{$B1$}\\
	\sigma_i\sigma_j & = \sigma_j\sigma_i~;~~~|i-j|>1,\tag{$B2$}
\end{align}
where $(B2)$ is the far commutativity of the braid group generators. The $s_i$ operators generate the permutation group,
\begin{align}
	s_is_{i+1}s_i & = s_{i+1}s_is_{i+1}, \tag{$S1$}\\
	s_is_j & = s_js_i~;~~~|i-j|>1,\tag{$S2$} \\
	s_i^2 & = \mathbb{I}. \tag{$S3$}
\end{align}
Each of the generators, $\sigma_i$ and $s_i$ act non--trivially on sites labelled by $i$ and $i+1$.
Furthermore, we have the {\it mixed} relations between the generators, $\sigma_i$ and $s_i$,
\begin{align}
	\sigma_is_j & = s_j\sigma_i~;~~~~|i-j|>1 \tag{$M1$} \\
	s_is_{i+1}\sigma_i & = \sigma_{i+1}s_is_{i+1}, \tag{$M2$} \\
	\sigma_i\sigma_{i+1}s_i & =  s_{i+1}\sigma_i\sigma_{i+1}. \tag{$M3$}
\end{align}
When all the above relations except $(M3)$ is satisfied the relations realize the virtual braid group, $VB_N$~\cite{virtual}. The relation $(M2)$ read backwards, $s_{i+1}s_i\sigma_{i+1}=\sigma_is_{i+1}s_i$ is always satisfied, however the relation $(M3)$ read backwards, 
\begin{equation}
	\sigma_{i+1}\sigma_is_{i+1}=s_i\sigma_{i+1}\sigma_i, \tag*{$(M3)'$}  
\end{equation}
is not always satisfied. If $(M3)'$ is satisfied instead of $(M3)$ we have the opposite loop braid group, $OLB_N$ and if both $(M3)$ and $(M3)'$ are satisfied we obtain the {\it symmetric loop braid group}, $SLB_N$. We will use the relations of $SLB_N$ along with 
\begin{equation}
	\sigma_is_{i+1}\sigma_i = \sigma_{i+1}s_i\sigma_{i+1}, \tag{$M4$}
\end{equation}
in constructing the solutions of the YBE.

Though the above algebraic relations of $SLB_N$ will suffice for the purpose of constructing $R$--matrices, we also mention the possibility of understanding the relations of the loop braid group pictorially. This is especially useful to visualize the exchange of loops realized by the generators of $SLB_N$~\cite{loop4}. 

\section{$R$--matrices from the loop braid group}
\label{sec:LBGRmatrices}
We now develop the technique to find $R$--matrices using $SLB_N$. The following {\it ansatz},
\begin{equation}\label{eq:Ransatz}
	R_{i, i+1}(u)\equiv R_i(u) = s_i + a(u)\,\sigma_i,
\end{equation}
is easily seen to satisfy the braided form of the YBE, 
\begin{equation}
	R_{i,}(u-v)R_{i+1}(u)R_{i}(v) = R_{i+1}(v)R_{i}(u)R_{i+1}(u-v),
\end{equation}
for an arbitrary function, $a(u)$, provided $s_i$ and $\sigma_i$ satisfy the relations of $SLB_N$ along with the relation in $(M4)$. Multiplying the $R$--matrix in Eq.~(\ref{eq:Ransatz}) with $s_i$ we obtain another $R$-matrix, 
\begin{equation}\label{eq:RSLB}
	s_iR_{i,i+1}(u) = \mathbb{I} + a(u)\,s_i\sigma_i,
\end{equation} that satisfies 
\begin{equation}
	R_{i, i+1}(u-v)R_{i, i+2}(u)R_{i+1, i+2}(v)=R_{i+1, i+2}(v)R_{i, i+2}(u)R_{i, i+1}(u-v)    
\end{equation}
which resembles the RTT relation with the $R$--matrices assuming the role of the monodromy matrices. As seen earlier this is the starting point for proving the integrability of various quantum models and the ABA. 

Having constructed the $R$--matrix out of the generators of $SLB_N$ we are left with finding appropriate representations. To this end we work on a closed chain comprising $N$ sites, with periodic boundary conditions. On each site we place a qubit with Hilbert space $\mathcal{H}_n =\mathbb{C}^2$ and the total Hilbert space is $\otimes_{n=1}^N~\mathbb{C}^2_n$. On this space the local representation for the permutation group generators, $s_i$ is well known,
\begin{equation}\label{eq:Permrep}
	s_i = \frac{\mathbb{I} + X_iX_{i+1}+Y_iY_{i+1}+Z_iZ_{i+1}}{2},
\end{equation}
where $X, Y, Z$ are the Pauli matrices,
\begin{equation}
	X = \begin{pmatrix*} 0 & 1 \\ 1 & 0\end{pmatrix*},~ Y=\begin{pmatrix*}[r] 0 & -\mathrm{i} \\ \mathrm{i} & 0\end{pmatrix*},~ Z=\begin{pmatrix*}[r]1 & 0 \\ 0 & -1\end{pmatrix*}.
\end{equation}
We are left with finding the representations of the braid group generators, $\sigma_i$, which we obtain from projectors that commute among themselves and are left invariant by the permutation operator. Such an operator can be written as,
\begin{equation}
	\label{eq:lbggenerator}
	\sigma_i = s_i + \alpha\,B_{i,i+1}.
\end{equation}
The operator $B_{i, i+1}$ satisfies
\begin{equation}
	\label{eq:Brelations}
	B^2_{i, i+1}  =  B_{i, i+1},~ B_{i, i+1}B_{i+1, i+2}=B_{i+1, i+2}B_{i, i+1},~s_iB_{i, i+1}=B_{i, i+1}.
\end{equation}
While there are plenty of projectors on $\mathbb{C}^2\otimes\mathbb{C}^2$ only a few among them are left invariant by an action of the permutation operator, $s_i$. One natural choice is when $B_{i, i+1}=P_iP_{i+1}$, that is an operator that is not `entangled' on the neighboring sites. As it is symmetric in the indices $i$ and $i+1$ it is left invariant under a left and right $s_i$ action. With these properties satisfied, a small computation shows that the operator in Eq.~(\ref{eq:lbggenerator}) satisfies the braid relations, with the inverse given by $\displaystyle \sigma_i^{-1} = s_i -\frac{\alpha}{1+\alpha}\,B_{i, i+1}$ and thus generates the braid group. A further calculation shows that the braid generator in Eq.~(\ref{eq:lbggenerator}) along with the permutation operator satisfies all the relations of $SLB_N$, especially the relations $(M3)$, $(M3)'$ and $(M4)$ in Sec.~\ref{sec:loopbraidgroup}.

Next we will use the permutation operator and the braid group generator in Eq.~(\ref{eq:lbggenerator}) to construct the $R$--matrix using the ansatz in Eq.~(\ref{eq:Ransatz}).

\section{Integrable spin chains}
\label{sec:LBGspinchains}

The Hamiltonian of the integrable system is taken to be a local operator following the discussion in Sec. \ref{subsec:XXXchain}. Using the formula for the local Hamiltonian in Eq.~(\ref{eq:localH}) with the $R$--matrix in Eq.~(\ref{eq:RSLB}), we obtain, 
\begin{equation}\label{eq:SLBH}
	H = \sum\limits_{i=1}^{N}~\frac{a'(u)}{1+a(u)}~\left[1+\frac{\alpha}{1+\left(\alpha+1\right)a(u)}~B_{i, i+1}\right],
\end{equation}
on a closed chain. According to the representation of the $B$ operators satisfying Eq.~(\ref{eq:Brelations}) we obtain both trivial and non--trivial spin chains. By trivial we mean those models which does not require the machinery of the ABA for a solution. The scope of the ansatz for the $R$--matrix in  Eq.~(\ref{eq:Ransatz}) is limited to this type of model but we would like to go further. 

In order to obtain a more non--trivial model, consider the ansatz,
\begin{equation}\label{eq:Ransatz2}
	R_{i, i+1}(u) = s_i + a(u)\,\sigma_i + b(u)\,\sigma_i^2,
\end{equation}
which is equivalent to working with,
\begin{equation}\label{eq:Ransatz3}
	R_{i, i+1}(u) = \mathbb{I} + a(u)\,s_i + b(u)\,B_{i, i+1},
\end{equation}
as we have the identities, 
\begin{equation}
	\sigma^{2k+1}_i = s_i + \left[\left(\alpha+1\right)^{2k+1}-1\right]\,B_{i, i+1},~~\sigma^{2k}_i=\mathbb{I}+ \left[\left(\alpha+1\right)^{2k}-1\right]\,B_{i, i+1}.
\end{equation}

The ansatz in Eq.~(\ref{eq:Ransatz3}) satisfies the YBE for $a(u)=\alpha u$, $b(u)=-2\alpha u$. Using Eq.~(\ref{eq:localH}), the resulting nearest neighbor term in the Hamiltonian is given by 
\begin{equation}\label{eq:intHB}
   H(u) = \sum\limits_{i=i}^N~\frac{1}{1-\alpha^2u^2}\left[(1-\alpha^2u)\mathbf{1} + \alpha(1-u) s_i + \left(2\alpha (u-1)\right)B_{i, i+1}\right].
\end{equation}
A naive reading of this local operator suggests that it is a deformation of the $XXX$--chain. While this is true for certain representations of $B_{i, i+1}$, there are many other non--trivial ones as we shall see below.


\paragraph{Model 1} - Consider the case when $B_{i, i+1} = P_iP_{i+1}$. A general projector on $\mathbb{C}^2$ is given by 
\begin{equation}
P=\frac{1}{2}\mathbf{1}+ l X + m Y+ n Z,
\end{equation}
provided $(l^2+m^2+n^2)=\frac{1}{4}$, where $l, m, n$ are complex numbers. This reduces Eq.~(\ref{eq:intHB}) to a deformation of the $XYZ$--spin chain.  
We include the cumbersome expression, 
\begin{align}
P_i P_{i+1} &=\frac{1}{4} \mathbf{1}+ l^2 X_i X_{i+1}+ m^2 Y_i Y_{i+1}+n^2 Z_i Z_{i+1} \notag\\ &+ \frac{l}{2} (X_i+ X_{i+1})+ \frac{m}{2}(Y_i +Y_{i+1})+\frac{n}{2} (Z_i +Z_{i+1})\notag \\&
+lm  (X_i Y_{i+1}+Y_i X_{i+1}) + mn (Y_i Z_{i+1}+Z_i Y_{i+1}) \notag \\  &+ln (Z_i X_{i+1}+X_i Z_{i+1}),
\end{align}
to give a glimpse of the models available for this choice of $B_{i, i+1}$. When $l=m =0$, we obtain a deformation of the $XXZ$-spin chain.


\paragraph{Model 2} - The $XXZ$--spin chain is obtained by choosing $\displaystyle B_{i, i+1} = \frac{\mathbb{I} + Z_iZ_{i+1}}{2}$ and the corresponding Hamiltonian is given by,
\begin{equation}
	H_{XXZ}\sim \frac{\alpha (1-u)}{2(1-\alpha^2 u^2)}[X_i X_{i+1}+ Y_i Y_{i+1} - Z_i Z_{i+1}].
\end{equation}
We note that the $XXZ$--spin chain is usually obtained from a {\it trignometric} $R$--matrix, but here we have shown a {\it rational} $R$--matrix (Eq.~(\ref{eq:Ransatz3})) to obtain the same model. 

\section{Outlook}
\label{sec:Outlook}
The usual approaches to obtaining integrable deformations of the $XXX$ and $XXZ$ chain include Drinfeld twisting of the associated quantum groups~\cite{kulish}. Here we have shown how the loop braid group can be used to obtain solutions of the YBE opening up the possibility of finding new integrable models. For a simple choice of the representation of the symmetric loop braid group we ended up with integrable deformations of the $XYZ$--spin chain. Using these $R$--matrices in the RTT relation we can find new $ABCD$ algebras. We intend to develop the algebraic Bethe ansatz method for such algebras in future publications. As a final remark, the fact that we obtain the $R$--matrices algebraically renders itself to an easy extension to higher spin chains.

\end{document}